# SUBMACROPULSE ELECTRON-BEAM DYNAMICS CORRELATED WITH HIGHER-ORDER MODES IN TESLA-TYPE SUPERCONDUCTING RF CAVITIES

A.H. Lumpkin*[1,3], R. Thurman-Keup, D. Edstrom, J. Ruan, N. Eddy, P. Prieto
[1]Fermi National Accelerator Laboratory, Batavia, IL 60510 USA
O. Napoly, [2]CEA-Saclay, Gif-sur-Yvette, France
B.E. Carlsten, K. Bishofberger
[3]Los Alamos National Laboratory, Los Alamos, NM 87545 USA

## ABSTRACT

We report the direct observations of sub-macropulse beam centroid oscillations correlated with higher order modes (HOMs) which were generated by off-axis electron beam steering in TESLA-type superconducting RF cavities. The experiments were performed at the Fermilab Accelerator Science and Technology (FAST) facility using its unique configuration of a photocathode rf gun injecting beam into two separated 9-cell cavities in series with corrector magnets and beam position monitors (BPMs) located before, between, and after them. Oscillations of ~100 kHz in the vertical plane and ~380 kHz in the horizontal plane with up to 600-μm amplitudes were observed in a 3-MHz micropulse repetition rate beam with charges of 100, 300, 500, and 1000 pC/b. However, the effects were much reduced at 100 pC/b. The measurements were based on HOM detector circuitry targeting the first and second dipole passbands, rf BPM bunch-by-bunch array data, imaging cameras, and a framing camera. Calculations reproduced the oscillation frequencies of the phenomena in the vertical case. In principle, these fundamental results may be scaled to cryomodule configurations of major accelerator facilities.

Index: 41.60

*lumpkin@fnal.gov.

# I. INTRODUCTION

Generation and preservation of bright electron beams are two of the challenges in the accelerator community given the inherent possibility of excitations of dipolar higher-order modes (HOMs) due to beam offsets in the accelerating cavities [1]. These long-range wake fields may deflect portions of the pulse train further off axis or lead to sub-macropulse beam oscillations. Such shifts on ultra-low emittance beams would have the potential for concomitant effects on averaged beam size and averaged emittance growth over a macropulse. Possible situations are related to the actual alignment of the cavities within the cryomodule or beam missteering into the cryomodules. The development of superconducting radiofrequency (SCRF) cavities for major facilities is exemplified by the L-band 9-cell TESLA-type cavity developed in the 1990's [2,3]. The cryomodule configuration with eight 9-cell cavities is currently the drive accelerator for the FLASH free-electron laser (FEL) [4], the European XFEL [5], the under-construction LCLS-II XFEL [6], the proposed MaRIE XFEL at Los Alamos [7], and the International Linear Collider (ILC) under consideration in Japan [8]. A recent study at FLASH using one specific $TE_{111}$ HOM showed that the root mean squared (rms) relative alignments were about 342 μm for the 40 cavities in the 5 cryomodules with some close to 600 μm off axis [9]. The assessment of the effects on beam quality of such implementations warrants further study as the thrust for brighter electron beams at higher powers continues.

In this article, we report on a unique opportunity to study HOM effects on beam parameters from two 9-cell TESLA-type cavities arranged in series after a photocathode (PC) rf gun in the electron linac at the Fermilab Accelerator Science and Technology (FAST) facility [10]. These studies focused on the correlated HOM effects on the beam itself, unlike several studies that used the HOMs to determine cavity field centers or cavity misalignments [9,11-14]. Moreover, there was the possibility for an additional ~10 m drift with all quadrupoles powered off prior to the dipole for the low-energy spectrometer. The upstream corrector magnets allowed controlled changes of the trajectories into the first capture cavity (CC1) which were monitored by the bunch-by-bunch capable rf beam position monitors (BPMs). In each cavity, there are two HOM couplers [e.g., see Fig. 2 of ref. 9] which are designed to outcouple and dampen many of the modes. The signals from these couplers are processed by four matched detector circuits to track the transverse modes in the first two dipole passbands of the cavities. We will describe the observed correlations between corrector current levels, HOM detector signals, and the beam average positions. Most importantly, we present conclusive evidence of the effect of these HOMs on the electron beam. We observed downstream transverse beam offsets that were linearly proportional to both the initial transverse offset and bunch charge. Moreover, we observed sub-macropulse 100-kHz and 380-kHz oscillations in beam position whose amplitude grew to the 600-μm level with the drift to the downstream BPM locations and which decayed with HOM Q values. We confirmed the variation of the offsets with detailed calculations which allowed us to infer which HOMs were being strongly excited. These effects were correlated with over a factor of 100 variation in HOM strength from selected charges and offsets. We also present a unique confirmation of the oscillation observed at the 25-μm level with a framing camera and the rf BPMs.

## II. EXPERIMENTAL TECHNIQUES

The FAST linac is based on an L-band rf photocathode (PC) gun which injects beam at a 3-MHz micropulse (or bunch (b)) repetition rate into two SCRF capture cavities denoted CC1 and CC2. The rf gun is a copy of the highly optimized PITZ gun [15], which has measured sub-μm emittances at 1 nC of micropulse charge. The gun includes a main solenoid for emittance compensation [16] and a bucking solenoid to ensure there is no residual magnetic field on the cathode. This is followed by transport to a low-energy electron spectrometer as shown in Fig. 1 with the beam properties for the studies summarized in Table 1. The total beam energy was 33 MeV with 4.5 MeV from the PC gun and about 14.2 MeV from each capture cavity. The range of micropulse charges is indicated with emittance compensation done as needed. A $Cs_2Te$ photocathode was irradiated by the UV component of the drive laser system described elsewhere [17]. The basic diagnostics for the HOM studies include the rf BPMs (denoted as B1xx) located before and between the two cavities as shown in Fig. 1, as well as the ten BPMs after the cavities and before the low energy spectrometer dipole. These are supplemented by the imaging screens inserted into beam line vacuum crosses (Xyyy) denoted at X107, X111, X121, and X124. The HOM couplers are located at the upstream and downstream ends of each SCRF cavity, and these signals are processed by the HOM detector circuits with the output provided online through ACNET, the Fermilab accelerator controls network. Recent upgrades included optimizing the HOM detectors' bandpass filters to target the two dipole passbands from 1.6-1.9 GHz and converting the rf BPM electronics to bunch-by-bunch capability with reduced noise [18].

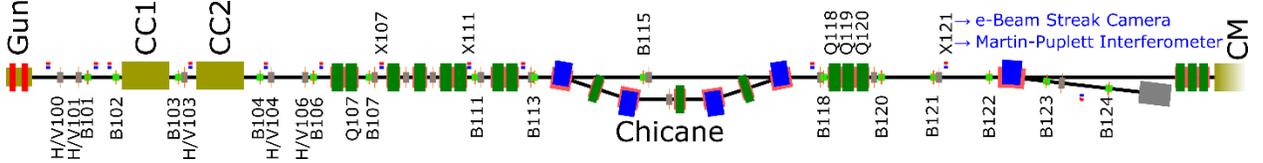

FIG. 1: Schematic of the FAST beamline showing the PC rf gun, CC1 and CC2 capture cavities, locations of the upstream correctors, rf BPMs, beamline components, and beginning of the cryomodule (CM).

A cryomodule with 250-MeV acceleration capability is located downstream, but it was not involved in these initial studies. An 858-kHz frequency detected in the BPM data was attributed to the CC2 low-level rf feedback system and removed in processing the centroid oscillations. We used 100- to 1000-shot averages of the BPM data for 1000 to 100 pC/b, respectively, to noticeably improve the statistical variances for the data.

Table 1: Summary of the FAST electron beam parameters used in the studies.

| Beam Parameter | Units | Value |
|---|---|---|
| Micropulse Charge | pC | 100 – 1000 |
| Micropulse Repetition Rate | MHz | 3 |
| Transverse Beam Sizes, (rms) | μm | 100 – 1200 |
| Transverse Emittance, normalized | mm-mrad | 1 – 5 |
| Bunch Length (rms) (compressed) | ps ps | 4 – 8 1 – 3 |
| Energy | MeV | 32 – 34 |
| PC Gun Gradient | MV/m | 40 – 45 |
| CC1 Gradient | MV/m | 12 – 15 |
| CC2 Gradient | MV/m | 12 – 15 |

### A. HOM Detectors

The four HOM detector circuits were matched in the electronics laboratory with each having the 1.3-GHz notch filter, the 1.6-1.9 GHz bandpass filter, the low-pass filter at 2.2 GHz, and the zero-bias Schottky detector. This configuration was chosen to cover the set of transverse dipole modes in the first two passbands of the cavities, particularly the M6 and M7 modes in passband 1 at 1.70 and 1.73 GHz and the M13 and M14 modes near 1.85 GHz in passband 2 which have the highest impedance as listed in a standard reference [19]. The other two filters in the HOM detectors reduce the 1.3 GHz fundamental by about 80 dB and reject the monopole modes above 2.2 GHz, respectively. We do miss the strong dipole mode in passband 3 near 2.58 GHz with this filter set, but the beam would still see this. Our strategy was to have our detectors targeting the modes that could give transverse kicks within the pulse train. The modes in the cavities at zero power were measured up to 3 GHz with a network analyzer with the cavities at 2 degrees K. The beam-driven modes were measured on an oscilloscope prior to the run, and the first two passbands were observed for each cavity.

### B. Beam Position Monitors

The rf BPMs are based on sets of four capacitive button pickups in the various vacuum crosses with matched cables out of the accelerator enclosure to the input of the electronics. The readout electronics were based upon architecture previously designed for the ATF damping ring at KEK [18]. Instead of downmix electronics in the tunnel, the FAST system uses new custom signal conditioning electronics [20] with bandpass filter cards at 325 MHz to ring the button signals for about 120 ns. The resulting 325 MHz signal is then under-sampled with the digitizer at 96MS/s locked to the 3-MHz bunch rate (32 samples/bunch). The same signal processing used at ATF for the turn-by-turn data is then applied to supply bunch-by-bunch information to the control system as device arrays which can be archived for each designated macro-pulse. At 500 pC/b, the position variances in the B101 and B102 data were 25-30 μm for individual bunches in a 50-micropulse train. The BPM resolution was dominated by the noise term so the 100-shot averages dramatically improved the effective spatial resolution to the few-micron level.

### C. Imaging Techniques

The beam-profile imaging stations predominately used vacuum crosses and hardware specified by FNAL staff and procured from RadiaBeam Technologies [21]. Two pneumatic actuators in series provided 4-position locations for a calibration target, YAG:Ce scintillator screen with 45 degree mirror behind it, an optical transition radiation (OTR) foil with 45 degree Al mirror behind it, and an impedance matching insert. The screens' emissions were imaged by a set of lenses and a 5-Mpixel Prosilica CCD camera with digital readout. The images were processed by online and offline toolkits.

We used one special configuration for imaging. The X121 station located 12 m downstream of CC2 had a configuration with the two screens in a holder positioned by a stepper actuator with the single OTR screen (Al-coated Si wafer) pointing beam left into an all-mirror optical transport to the framing camera and the YAG:Ce screen/mirror pointing beam right for optical transport to a standard Prosilica CCD camera. The framing camera was based on a Hamamatsu C5680 mainframe operating with the M5677 low-speed vertical unit and the M5679 dual-time-axis horizontal sweep unit. The readout camera was a Prosilica 1.3-Mpixel CCD digital camera. The selectable deflection rates of the two units allowed display of an array of individual micropulse images within one readout image. The technique was pretested with the 3-MHz green component of the drive laser and X121 OTR signals previously [22]. In the present case, we imaged 18 consecutive micropulses, and we adjusted trigger delays to the deflection circuits to select which of the 50 micropulses in a macropulse were displayed subject to the internal delays of the units.

### III. EXPERIMENTAL RESULTS

The effects of the HOMs on the transverse beam parameters were investigated by measuring time-resolved centroids with both BPMs and a framing camera. In the former case, we had the most success with the rf BPM bunch-by-bunch data averaged over 100-1000 macropulses depending on the charge per micropulse, or bunch. We will start discussions with the HOM results.

#### A. HOM Results

Examples of the digitized detector output waveforms from both the upstream and downstream CC1 detectors for a 50-bunch train at 3 MHz, 500 pC/b, and a V101 corrector current at 1A are shown in Fig. 2a. The signals rise rapidly in the first 10 bunches, and then stabilize at a peak value. In Fig. 2b, we show the sum of the peak values for the four detectors as a function of the upstream vertical corrector current. This shows that the HOM sums are minimized near V101=0.0 A while increasing by almost an order of magnitude near V101= ±1A for this charge of 500 pC/b. For a fixed corrector setting, the HOM signals do vary linearly with charge from 100-1000 pC/b as expected as shown in Fig. 2c for the CC1 detector 1 (upstream) case. In the case of H101 scans, we found the HOM signal values were reduced at H101= 0.42 A, and the CC2 HOM signals were further reduced by using H103= -0.82 A and V103= -0.37 A, which could be explained by some intercavity misalignments.

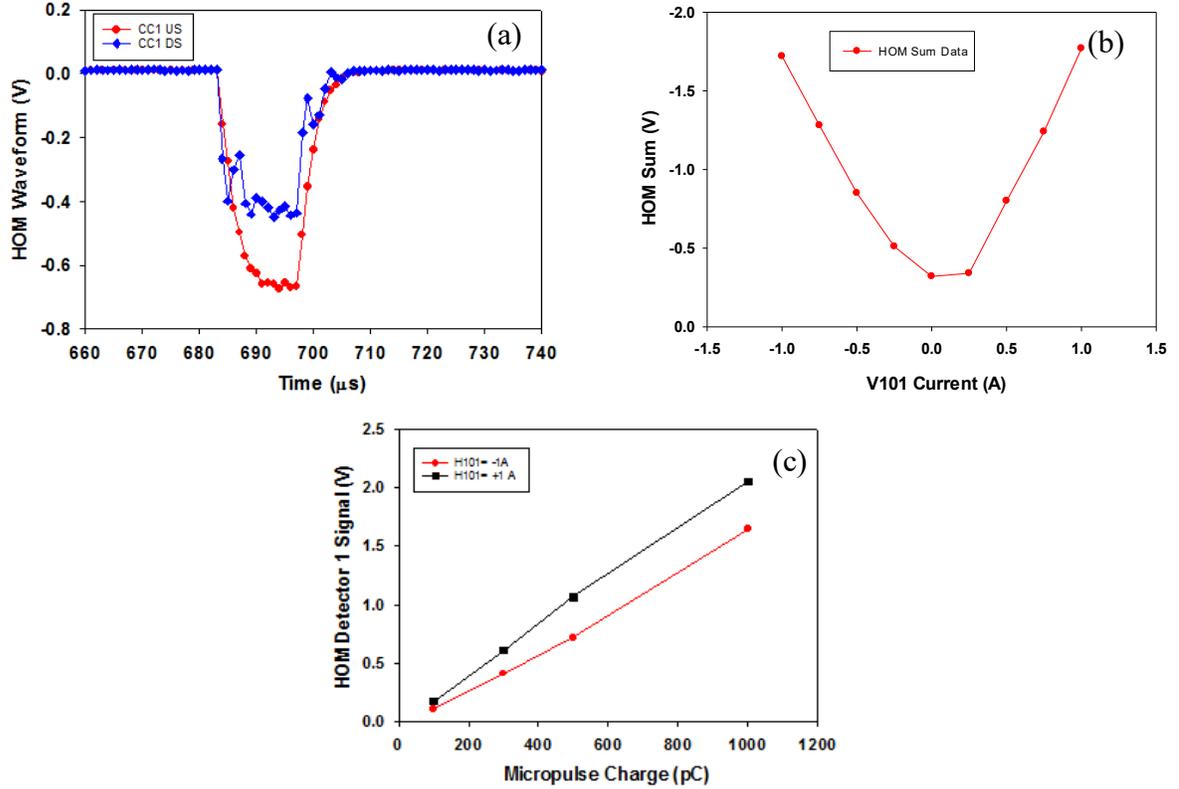

FIG. 2: a) Examples of the CC1 upstream and downstream HOM detector waveforms for 50 b at V101=1 A b) Measured sums of the 4 HOM detector signal peak values as a function of the vertical corrector current (V101) for 500 pC/b and c) plot of the near linear dependence on Q of CC1 HOM detector 1 for H101= -1 and +1 A from the reference current.

### B. Beam Position Monitor Results

During the V101 vertical corrector current scan, the average vertical beam positions into and out of the cavities were tracked as shown in Fig. 3. Although B101 showed a slight effect since it is located after V101, B102 located 0.5 m downstream shows significant position changes of +-5 mm at -+1 A. These implied up to ~10-mrad kick angles into CC1 and 15 mm offsets at the extremes of the corrector currents used. However, the B103 readings for the -1 A case indicated about a 6-mm offset entering into CC2 after accounting for B103's 4-mm mechanical offset from the beamline axis. This offset difference of 15 mm to 6 mm is attributed to cavity focusing effects in CC1 as seen in Fig. 3b. The lines are the extrapolated trajectories through CC1 and CC2 using the transfer matrix of Chambers [23] with B101 and B102 forming the initial vector. The trajectories after CC2 were adjusted with the correctors H/V104 and H/V106 to maintain the beam position at B106 and BPMs located further downstream.

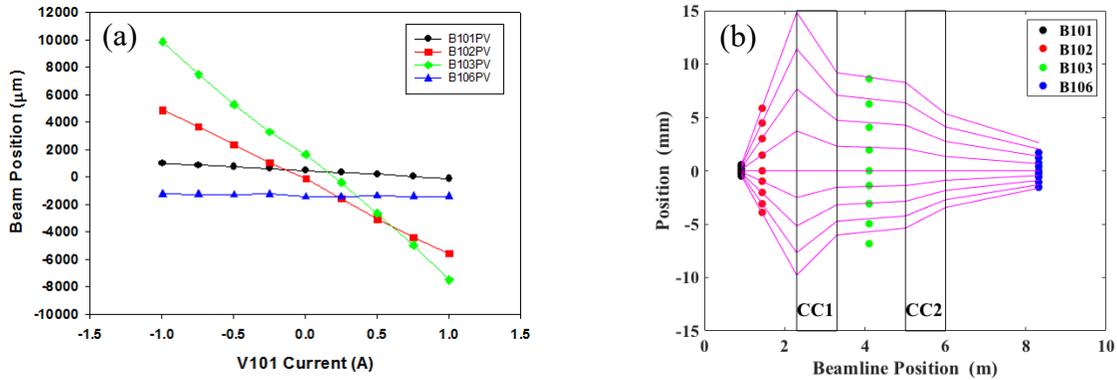

FIG. 3: (a) Macropulse-averaged vertical beam positions tracked with rf BPMs before, between, and after the two capture cavities versus the V101 corrector settings. (b) Beam positions relative to the 0-A case as a function of BPM position for a sequence of V101 corrector settings.

The clearest indications of the centroid shifts were found in the bunch-by-bunch rf BPM data with 100-shot averages. In these cases, the averaged beam position of the pulse train was subtracted from each of the bunch positions for display purposes. The averaging proved to be a critical step and relied on the reproducibility of the beam and the excited HOMs over the 100 shots at 1 Hz. These striking effects were seen in the plane of the beam steering by V101 and correlated with the direction. The vertical BPMs showed the strongest effects with a 100-kHz oscillation in the first

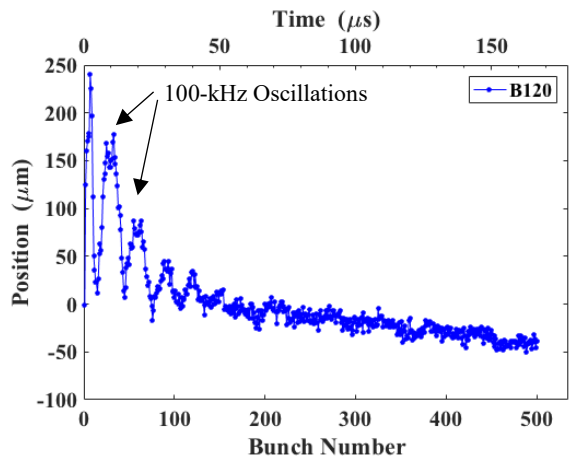

FIG. 4: Vertical centroid oscillations shown at rf BPM location B120 for 500 b, 500 pC/b, and V101=+1A. The 100-kHz oscillation decays noticeably in the first 200 b, and a centroid slew continues to the end of the macropulse.

200 b whose amplitude increased with drift from CC2 down the beam line. In Fig. 4, we show the effects at B120 which is 10 m downstream of CC2. In this case, the centroid slew continues to the end of the macropulse.

We focused on the first 50 b with different corrector scans. In these data, the upstream corrector values and H/V103 were set for HOM detector signal minima in CC1 and CC2 at 500 pC/b. Examples for 50-b pulse trains are shown in Fig. 5 for B107 which is ~3.3 m after CC2, B113 which is 5.5 m downstream of CC2, and B120 and B121 which are 10.5 and 11.5 m downstream of CC2, respectively. The sets for V101 = -1, 0, +1 A and 1000 pC/b are shown. Note the oscillation changes phase/direction with V101 polarity and HOM direction. At this higher charge, the HOM-induced oscillations are still visible for V101=0 A. This is probably due to an imprecision in finding a minimum for both CC2 HOM detector signals at the same time. The error bars are the same as the size of the data points. The nominal HOM sum signals corresponding to these corrector values were shown in Fig. 2b.

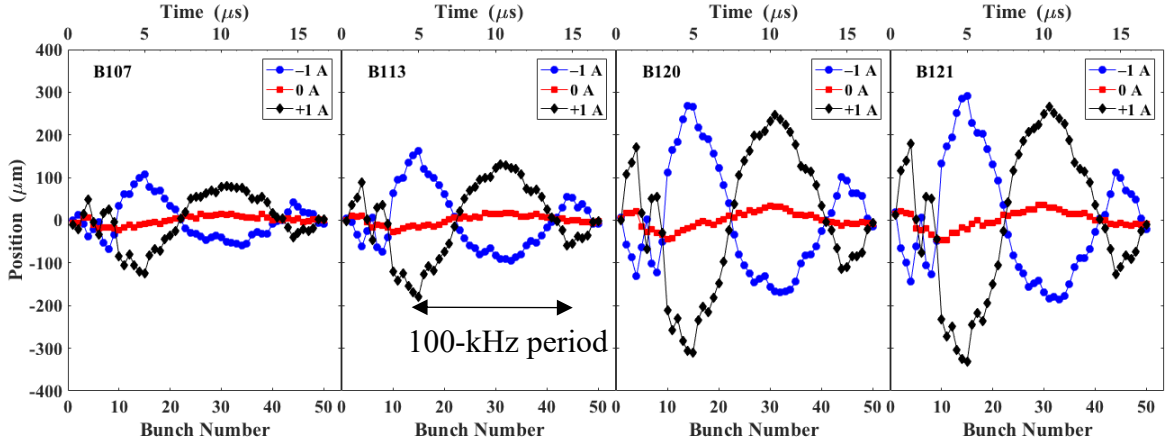

FIG. 5: Centroid vertical oscillations at 100 kHz shown at rf BPM locations B107, B113, B120, and B121 downstream of CC2. The three curves are for corrector currents at V101= -1, 0, +1 A, respectively. A strong correlation of sub-macropulse beam motion with V101 current was seen, consistent with excited HOMs as in Fig. 2.

As further investigation in a subsequent run and with the correctors set to minimize the HOMs as a reference point for V101= 0.0 A, we transported the beam to the low energy spectrometer with all the quadrupole currents at zero. This allowed the ten BPMs along the beam line to record the growing oscillation amplitude with drift from the postulated original kick angle generated by the HOMs. The 100-kHz oscillation for the 3-MHz beam was matched by the CC2 mode M14 in the second passband with a specific polarization angle to be discussed subsequently. The envelope of the centroid oscillation was tracked by evaluating the extremes of the oscillation as shown in Fig. 6a. The angles of the extrema were projected back through CC2 and CC1 using a matrix propagation including the cavity matrix of Chambers [23] as seen in Fig. 6b. These data are for the 1000 pC/b case with the nominal corrector current scan set of -1, 0, +1 A in the V101 corrector.

The averaged array beam position was subtracted for each bunch position. The data indicated a maximum kick angle of 84 µrad corresponding to the peak amplitudes of the oscillations. The deduced maximum kick angles for 100, 300, and 500 pC/b were 8, 25, and 42 µrad, respectively. The results are consistent with the HOMs varying linearly with charge.

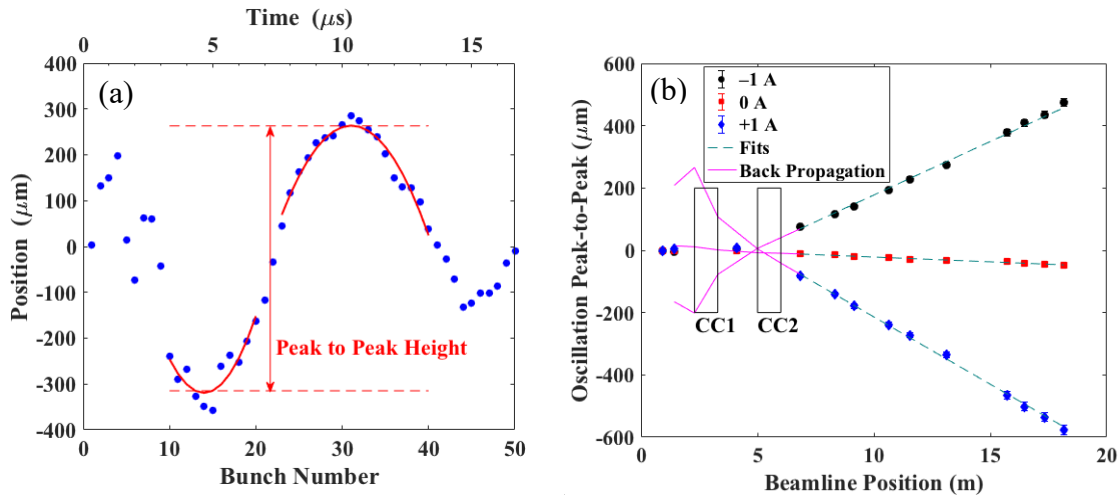

Figure 6: Analysis of the V101 scan for the 1000 pC/b with 50-bunch data is shown: a) identification of the 100-kHz oscillation (30-bunch period at 3 MHz) peak to peak amplitude within the 50-bunch train and b) the fitted amplitude as a function of drift in meters down the beamline as measured by the ten BPMs. The plot also shows the schematic positions of CC1 and CC2 cavities as the rectangles with the back propagated oscillation through CC2 and CC1. The oscillation likely originated in CC2 since the B103 data do not match the back-propagation line.

We next performed corresponding tests with the H101 horizontal corrector also located 0.5 m before B102 and the CC1 cavity. In this case the minimization of HOMs in CC1 had required a corrector current value of 0.42 A as found by scanning the corrector value and tracking the HOMs in CC1 and CC2. Therefore, the scan was 1 A from this reference value in each direction. In this case we found a different oscillation frequency of 380 kHz plus a mixture of another mode. The shape of the centroid motion is seen in the 1000 pC/b case of Fig. 7a. The envelope of the centroid oscillation was tracked by evaluating the extremes of the oscillation as also shown in Fig. 7a. As with the previous measurement, the angles of the extrema were propagated back through CC2 and CC1 as seen in Fig. 7b. In this case we show data from 5 different corrector settings in 0.5-A steps from the reference value. Here the B103 monitor located between the cavities indicated a non-zero oscillation amplitude consistent with the back propagation and implying the source of the oscillation was prior to it (i.e., originating in CC1). This centroid movement was different than that in Fig. 6 which one can now see also involved an offset and slew in centroid which was

11attributed to a combination of two $TM_{110}$ modes in CC1. The H101= +1A data indicated a maximum kick angle of 37 µrad corresponding to the peak amplitudes of the oscillations at B122. The deduced kick angles for 100, 300, and 500 pC/b were 3.5, 11, and 18.5 µrad, respectively as shown in Fig. 7c. As with the vertical data, these results are consistent with the HOMs' varying linearly with charge.

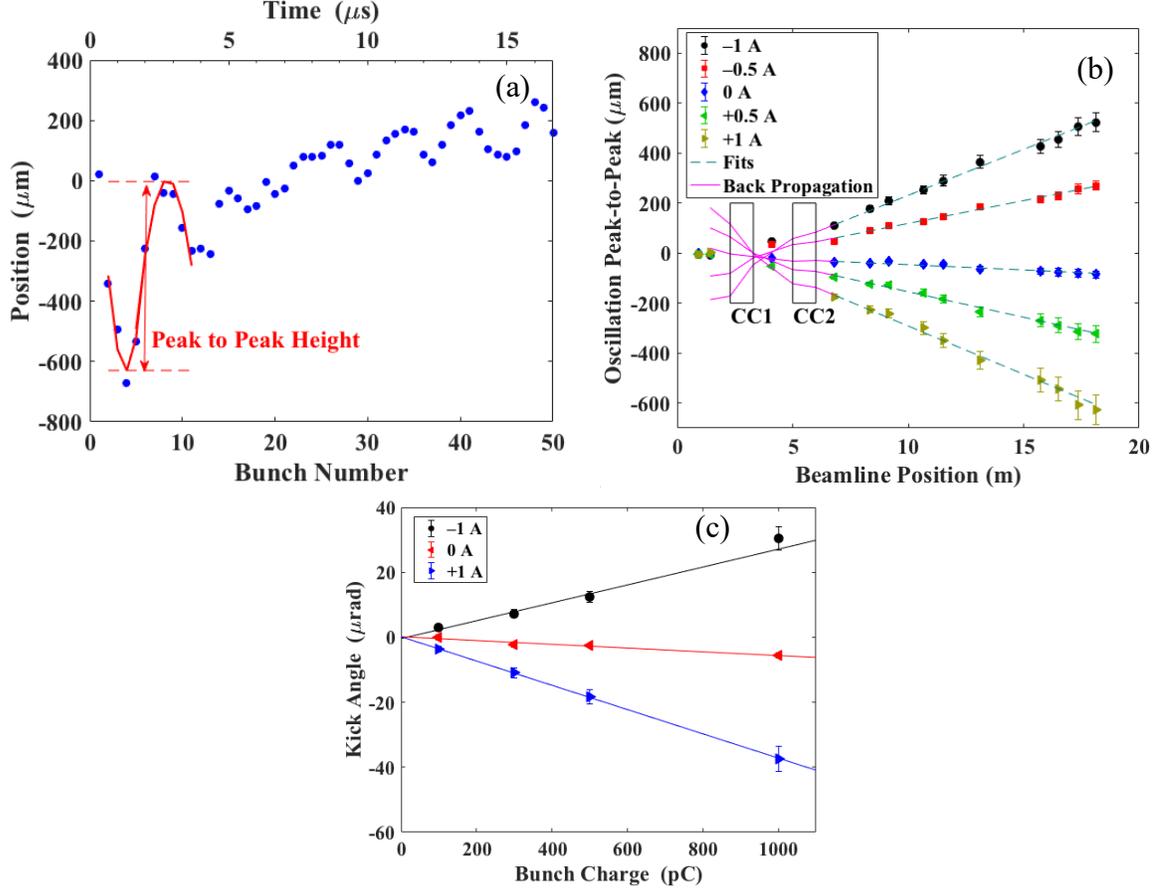

Figure 7: Analysis of the H101 scan for the 1000 pC/b with 50-bunch data is shown: a) identification of the 380-kHz oscillation (8-bunch period at 3 MHz) peak to peak amplitude within the 50-bunch train and b) the fitted amplitude as a function of drift in meters down the beamline as measured by the ten BPMs for five corrector settings. The schematic positions of CC1 and CC2 cavities are shown as the rectangles with the back propagation being consistent with B103 data, implying the source was in CC1, and c) plot of the linear variation of CC1 HOM kick angle with charge for the H101= -1, 0, +1 A cases.

## C. Framing Camera Results with X121 OTR

We made an important confirmation of the unexpected bunch-by-bunch 100-kHz centroid oscillation with a completely independent measurement using a unique framing-camera technique. This camera was capable of a resolution of less than 6 µm at the X121 YAG:Ce screen and the OTR screen location for a beam charge of 250 pC/b. To facilitate the transport and preserve image intensity for a single micropulse in the camera, we adjusted the upstream quadrupole triplet currents at Q118-120 to produce a beam focus of ~200 µm at the X121 location. However, this step reduced the magnitude of the bunch-by-bunch centroid oscillations from ~ 500 µm to about 30 µm as shown in Fig. 8. We successfully imaged 18 consecutive micropulses with 1000 pC/b at 3 MHz in one 5-µs vertical sweep as shown in Fig. 9a. Each one of the 18 images' projected y profiles was fitted to a single Gaussian peak, and the amplitudes, centroids, and sigmas with errors were determined. In this case the X121 image is rotated 90 degrees in the optical transport to the streak camera entrance so the x-display axis is the original y axis. We used a trigger delay to select bunches 11 to 28 and 31 to 48 in separate framing camera images which together would cover the 100-kHz vertical oscillation cycle from bunch number 15 to bunch number 45. Using the calibration factor of 6.6 µm/pixel for the framing camera, an oscillation amplitude effect of ~4 pixels or ~25 µm was seen (consistent with the rf BPM data also shown). The 3-image sum positions for the +1 A and -1 A corrector settings were evaluated with the 0-A data column used as reference. The 100-kHz cycle is discernable (Compare to Fig. 8's B121 oscillation period) as well as the good agreement with the B121 data in the combined bunch-by-bunch centroid plot shown in Fig. 9b. Image averaging and higher charges could further reduce statistical fluctuations. The technique could also be applied at higher micropulse repetition rates such as planned for MaRIE where BPMs may not achieve reliable bunch-by-bunch performance.

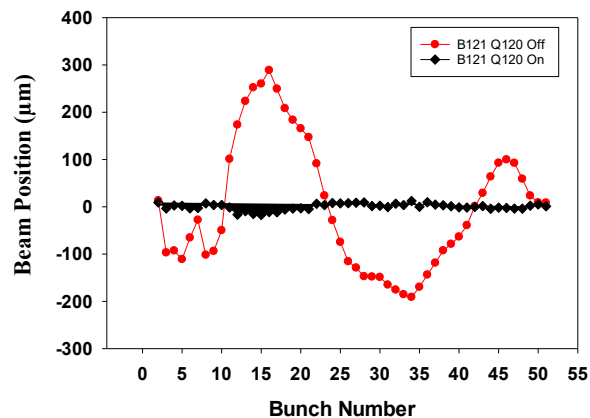

Figure 8: Comparison of the 50-bunch centroid oscillations at B121 located after the Q118-120 quadrupole triplet with V101= -1A. The downstream B121 location shows the oscillation amplitude was much reduced to about 30 µm in the Q120 powered ON case. The beam actually appears to be over focused since the deflection phase is the opposite of the Q120 OFF case as confirmed in Fig. 9.

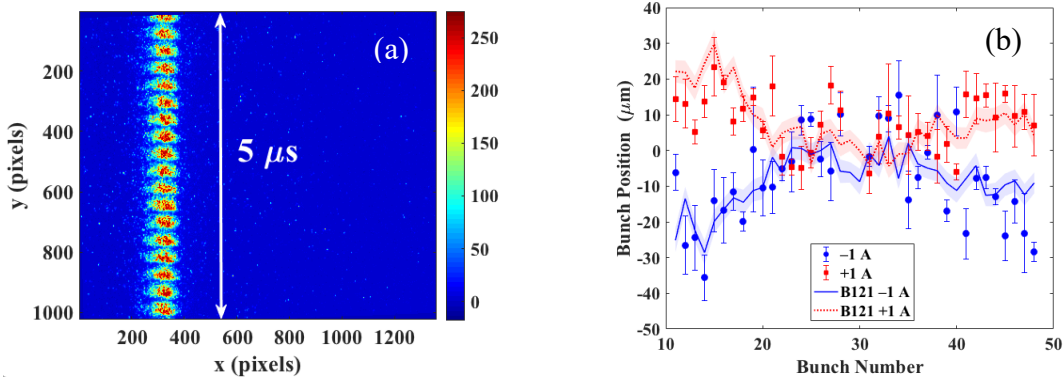

Figure 9: Examples of a) imaging 18 consecutive micropulses (31-48 out of 50) using the X121 OTR screen and the framing camera mode of the C5680 streak camera with V101=+1 A and b) The bunch-by-bunch centroids for V101=-1 and +1A referenced to the V101= 0A case for 36 of 50 bunches with *direct* comparisons to the X121 BPM data. This camera setup used a 5-µs vertical sweep and 100-µs horizontal sweep range. The CCD image was rotated 2 degrees counter clockwise to mitigate an instrumental effect from the horizontal sweep for display. The 100-kHz centroid oscillation is discernable, and the camera data and the BPM data are in good agreement.

## IV. ANALYSIS OF RESULTS

Theoretical calculations of the effects of the different modes were implemented as an evaluation of the kick angle by the beam-driven HOMs. The angular kick $\delta \vec{r}'(s)$ experienced by a trailing electron of charge $e$, velocity $\vec{v}$, and momentum $\vec{p}$ at a distance $s$ from the HOM-exciting bunch of charge $Q_b$ and transverse offset $\vec{r}_0$ is given by

$$\delta \vec{r}'(s) = \frac{\Delta \vec{p}_\perp(s)}{p} = \frac{e}{pc} \int \left( \vec{E}_\perp + \vec{v} \times \vec{B} \right)(s) \cdot dl = \frac{e}{pc} Q_b \vec{W}_\perp(s), \quad (1)$$

where $\vec{B}$ and $\vec{E}_\perp$ are the magnetic and transverse electric fields generated by the HOM-exciting bunch, the integral is over the electron path, and $c$ is the speed of light. For a series of $m$ bunches, the wake potential at the $m^{\text{th}}$ bunch, $\vec{W}_\perp(s_m)$, is given by the following summations over the resonant dipole modes, $n$, and the previous bunches, $k$:

$$\vec{W}_\perp(s_m) = \vec{r}_0 \frac{c}{2} \sum_{k=1}^{m-1} \sum_n \left( \frac{R_\perp}{Q} \right)_n e^{-\frac{\omega_n^2 \sigma_z^2}{2c^2}} \sin\left( \frac{\omega_n(s_m - s_k)}{c} \right) e^{-\frac{\omega_n(s_m - s_k)}{2Q_n c}} \cos^2(\varphi_n), \quad (2)$$

where $\omega_n$ is the angular frequency, $\left(\frac{R_\perp}{Q}\right)_n$ the transverse impedance, $\varphi_n$ the polarization angle, and $Q_n$ the damping factor of mode $n$. For the low-frequency passbands under investigation, the first exponential factor is equal to 1 since the mode wavelengths are much larger than the bunch length $\sigma_z$. A network analyzer was connected to CC1 and CC2 while cooled at 2K to measure the angular frequencies and damping factors, and the transverse impedances were calculated in [18], but the polarization angles are unknown and could be measured in cold state only through beam-based measurement. The beam-excited HOM spectrum was measured in CC2 along with the calculated shunt impedances for each mode and found to be consistent with Reference [19]. As expected, modes with higher shunt impedances were rung to higher amplitudes.

The 100-kHz and 380-kHz oscillations seen in the data suggest a beating between a dominant HOM and the beam frequency, and we can qualitatively reproduce the oscillatory data (with even some quantitative agreement) using the assumption of one or two dominant modes. Using the cold frequency measurements, we observed that the difference in frequency between one of mode 14's polarizations in CC2 (M14, at 1.88 GHz) and the bunch frequency harmonic was 100 kHz. This, together with the implication of CC2 in Fig. 6, leads us to conclude that this mode was responsible for the vertical centroid oscillations. A calculation of the angular kicks from M14 is shown in Fig. 10a. The modeling is mode polarization dependent, but shows oscillatory effects at 100 kHz in some cases in CC2 for the mode M14 with polarization angle vertical. These calculations assumed a translation of the beam off axis by 5 mm at 500 pC/b instead of our simplistic experimental approach of single corrector current changes resulting in kick angles of 0-10 mrad into CC1. The amplitudes of the oscillations were underestimated in the vertical case compared to the data, and this aspect remains a topic for further study.

The effect of the bunch summation in Eq. 2 is to cause oscillations due to the phase difference in contributions from different bunches. However, due to the various mode Qs, the second exponential term forces contributions to vanish from bunches preceding the observer bunch by more than some Q-dependent time. For M14, this decay time is 100 μsec or more and thus the effect reaches saturation for bunch numbers greater than 100, consistent with the results seen in Fig. 4.

The inferred solution for the horizontal steering and HOM excitation is for two modes in CC1 that are summed to give a slew in position with an oscillation overlaid as shown in Fig. 10b.

These are M7 at 1.73 GHz and M30 in passband 3 at 2.58 GHz [16] with a maximum kick angle of ~22 μrad. Comparing to the data in Fig.7, this estimate would be scaled up by a factor of two to an offset kick of 44 μrad assuming an average linear offset of 10 mm (since the beam was projected to enter CC1 15 mm off axis and measured to exit 6 mm off axis). There would be another factor of 2 increase for the 1000 pC/b case. The observed 380 kHz frequency in the data was not matched in this case, but the trends are quite similar.

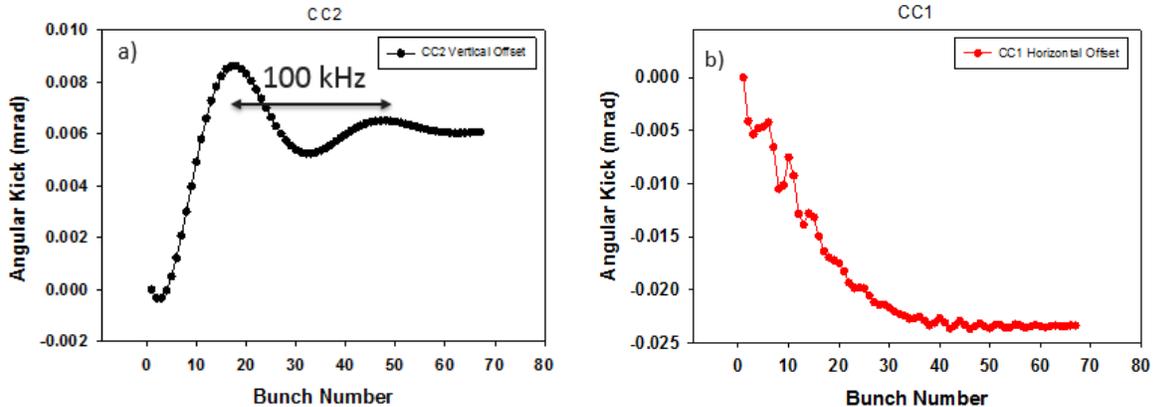

Figure 10: Calculational results for 500 pC/b and a 5-mm beam offset entering CC1 and CC2 a) the vertical steering and vertical oscillation for CC2 (M14) generated dipole HOMs and b) the horizontal steering and horizontal position oscillations for CC1 (M7 and M30) generated dipole HOMs.

## V. SUMMARY

In summary, we have elucidated the clear correlation of HOMs and sub-macropulse centroid motion with 100-kHz frequency for vertical steering and 380 kHz for horizontal steering of a 3-MHz pulse train injected off-axis into TESLA-type cavities. Oscillation amplitudes were observed up to 600 μm in the BPMs depending on the drift and lattice settings at FAST, and oscillation amplitudes of ~25 μm were observed for the first time with a framing camera technique. Since we had used rf feed-forward techniques to flatten the rf field amplitudes during the macropulse time in CC1 and CC2, we noted the measured waveforms did not account for, nor correlate with, our observed sub-macropulse effects. The former such centroid motions would lead to macropulse-averaged beam size and emittance growth downstream of the two cavities, but this is particularly relevant for ultra-low emittances in the 0.1-mm mrad regime in other facilities. The comparisons of experiment and calculational results are part of a benchmarking effort for such TESLA-type cavities. We anticipate applying these techniques to the FAST cryomodule in the next year. We suggest these data could also be useful in benchmarking start-to-end simulations of beam quality preservation over many cryomodules. In principle, these results are relevant to the specifications of cavity alignments in cryomodules, HOM coupler design, and cavity symmetry, as well as the requirements for beam trajectories into SCRF cavities to preserve ultra-low transverse emittances in the major accelerator facilities both existing and proposed.

## ACKNOWLEDGEMENTS

The authors acknowledge the technical support of J. Santucci, D. Crawford, and B. Fellenz; the project support of J. Liebfritz; the mechanical support of C. Baffes; the lattice assistance of S. Romanov; the cold cavity HOM measurements of A. Lunin and T. Khabiboulline of the Technical Division, the SCRF support of E. Harms; discussions with S. Yakovlev and Y. Shin; as well as the discussions with and/or support of A.Valishev, D. Broemmelsiek, V. Shiltsev, and S. Nagaitsev of the Accelerator Division at Fermilab. The Fermilab authors acknowledge the support of Fermi Research Alliance, LLC under Contract No. DE-AC02- 07CH11359 with the United States Department of Energy. The Los Alamos authors gratefully acknowledge the support of the US Department of Energy through the LANL/LDRD Program for this work.